\documentstyle[aps,prl,multicol,epsf]{revtex}
\draft
\begin{document} 
\title{Mesoscopic fluctuations of the Coulomb drag at $\nu$=1/2}
\author{B. N. Narozhny and I. L. Aleiner}
\address{Department of Physics and Astronomy, SUNY at Stony 
Brook, Stony Brook, NY 11794}
\author{Ady Stern}
\address{Department of Condensed Matter Physics, Weizmann
Institute of Science, Rehovot 76100, Israel}
\maketitle

\begin{abstract}

We consider mesoscopic fluctuations of Coulomb drag transresistivity
between two layers at a Landau level filling factor $\nu=1/2$ each.
We find that at low temperature sample to sample fluctuations exceed
both the ensemble average and the corresponding fluctuations at $B=0$.
At the experimentally relevant temperatures, the variance of the
transresistivity is proportional to $T^{-1/2}$. We find the
dependence of this variance on density and magnetic field to reflect
the attachment of two flux quanta to each electron.
\end{abstract}

\date{Draft: \today}
\begin{multicols}{2}

Measurements of Coulomb drag between two parallel electronic systems 
at close proximity \cite{exp1,exp2,exp3,exp5,exp6} are a useful 
tool for studies of electron-electron interactions. In these 
measurements a current $I_1$ is driven in one ("active") of the systems. 
As a consequence of inter-layer electron-electron scattering, a 
momentum $q$ is transferred from the active system to the other 
("passive") one with a voltage $V_2$ developing in the latter. The 
ratio $\rho_D=-V_2/I_1$ is known as the transresistivity or the drag 
coefficient.  

For clean systems at zero magnetic field\cite{the1,the2,the3} 
$\rho_D\propto T^2$ (with $T$ being the temperature), as follows from 
Landau's phase space argument. Recently, Coulomb drag was studied 
experimentally \cite{exp5,exp6} in a system of two two-dimensional 
electron layers in a strong magnetic field $B$. When the Landau level 
filling factor in each layer is $\nu=1/2$, the effect is enhanced by 
several orders of magnitude relative to $B=0$, and follows a 
sub-quadratic temperature dependence. Moreover, at very low 
temperature it was found to saturate to a non-zero value. Theoretically 
\cite{the4,the5}, $\rho_D\propto T^{4/3}$ with a prefactor much larger
than at $B=0$. In the presence of disorder\cite{the3,the6}, both at 
$B=0$ and at $\nu=1/2$, $\rho_D\propto T^2\ln T$. 

Mesoscopic fluctuations of Coulomb drag 
at $B=0$ were recently studied
theoretically\cite{na}. When $k_F d\gg 1$ ($k_F$ is the Fermi momentum
and $d$ the inter-layer separation), Coulomb drag is proportional to
$\left(\frac{\partial\sigma}{\partial n}\right)^2$, where $\sigma$ is
the single layer conductivity and $n$ is the electron density. On
average, the derivative is $\sigma/n$. Its fluctuations for a fully
coherent sample are $e^2/\hbar E_T N(0)$ with $E_T\equiv\frac{\hbar
D}{L^2}$ being the Thouless energy and $N(0)$ being the thermodynamic
compressibility of electrons. Hereafter $D$ is the diffusion constant
of electrons.  In the absence of phase breaking, for large enough
samples the fluctuations exceed the average: $\langle
\rho_D^2\rangle/\langle \rho_D\rangle^2 \simeq (L/l)^4$, where $l$ is
the elastic mean free path.  As usual, mesoscopic fluctuations of the
drag result from quantum interference between different trajectories
and thus are suppressed by phase breaking processes, leading to a
non-trivial temperature dependence.

In this paper we study fluctuations of Coulomb drag in the $\nu=1/2$
case, where each layer is described\cite{hlr} as a Fermi liquid of
composite fermions. Each composite fermion is an electron with two
attached flux quanta \cite{Jain}, that interacts with the others both
electrostatically and by means of a Chern-Simons interaction. While
most properties of the $\nu=1/2$ state can be analyzed in terms of
semi-classical dynamics of composite fermions, this study requires a
quantum mechanical analysis.

We start with a qualitative discussion, then sketch the calculation
whose details are to be published elsewhere, and conclude by
suggesting experimental probes to distinguish the effect of mesoscopic
fluctuations from other phenomena. All along, we use $\sigma$ and
$\rho$ to denote the measurable electronic conductivity and
resistivity. Composite fermion conductivity, resistivity, and
diffusion constant are denoted by $\sigma^{cf}$, $\rho^{cf}$, and 
$D^{cf}$. For
simplicity we consider only the longitudinal drag conductivity and
resistivity, denoted by $\sigma_D$ and $\rho_D$. We note in
passing that the Hall drag resistivity, which vanishes when
disorder-averaged, would fluctuate with the same correlation functions
as the longitudinal one, $\rho_D$.

It is convenient to calculate mesoscopic fluctuations for the drag 
conductivity, $\sigma_D$, rather than for $\rho_D$. In fact, at 
$\nu=1/2$, to the lowest non-vanishing order in the screened interlayer 
interaction \cite{the4}, 

\begin{equation}
\rho_D \approx \left(\frac{2h}{e^2}\right)^2 \sigma_D,
\label{rn}
\end{equation}

\noindent
In the same approximation, $\sigma_D$ can be expressed in terms of 
the non-linear susceptibility of the system $\Gamma$ (to be defined 
below) and the propagator ${\cal D}$ of the screened interlayer Coulomb 
interaction: 

\begin{equation}
\sigma_D= {{1}\over{4 S}} 
\int{{d\omega}\over{2\pi\hbar}}
\left({{\partial}\over{\partial\omega}} 
\coth{{{\hbar\omega}\over{2T}}}\right)
{\cal D}^R_{12}\Gamma^{x}_{23}
{\cal D}^A_{34}\Gamma^{x\; *}_{41},
\label{dia}
\end{equation}

\noindent
Here, $S$ is the area of the sample,
numerical subscripts indicate spatial coordinates, 
and are implied to be integrated over. The non-linear susceptibility 
of electrons $\Gamma_{ij}^\alpha(\omega)$ is a response function 
relating a voltage $V(r_i)e^{i\omega t}$ to a $dc$ current it induces 
by the quadratic response:

\begin{equation}
J^\alpha=\int d{\bf r}_1\int d{\bf r}_2\Gamma_{12}^\alpha(\omega) 
V({\bf r}_1)V({\bf r}_2),
\label{gamma-el}
\end {equation}

\noindent
with $\vec{J}$ being the induced $dc$ current. From  
gauge invariance $\int d{\bf r}_1\Gamma_{12}^\alpha(\omega) = 
\int d{\bf r}_2\Gamma_{12}^\alpha(\omega) = 0$.

The retarded (advanced) propagators of the screened interlayer 
interaction ${\cal D}^{R(A)}$ are calculated by means of a standard 
random phase approximation (see, e.g. Ref.~\onlinecite{ady}).
In momentum representation,
\begin{equation}
{\cal D}^{R(A)}=\frac{\left[\Pi^{R(A)}\right]^{-1}}{1+2\pi e^2d\Pi^{R(A)}},  
\ \ \Pi^{R(A)}=\frac{N(0)Dq^2}{\mp i\omega+Dq^2}.
\label{Ds}
\end{equation}

\noindent
In the diffusive regime the non-linear susceptibility $\Gamma$ can be 
obtained from  Ohm law,  
%\begin{equation}
$
\vec{j} = {\hat\sigma }\vec{E} - e D {\bf \vec\nabla} n,
%\label{o}
%\end{equation}
$
%\noindent
where $\vec E$ is the electric field. Combined with the continuity 
equation,  Ohm law yields for the linear response of the density 
to an applied field,  

\begin{equation}
\langle n(\vec{q},\omega)\rangle = \frac{1}{e} 
\frac{iq^\alpha\sigma^{\alpha\beta}E^\beta(\vec{q},\omega)}
{-i\omega + D q^2}.
\label{den}
\end{equation}

\noindent
where $\langle\dots\rangle$ indicate disorder-averaged quantities.

A non-linear response to the electric field results from the density 
dependence of the conductivity $J_{dc} = {\rm Re} \left(
\frac{\partial \sigma}
{\partial n}\right) n(q,\omega) E(-q,-\omega)
$, and yields, 

\begin{equation}
\langle\Gamma^\gamma\rangle = 
\frac{1}{e} 
\frac{\partial\langle\sigma^{\gamma\delta}\rangle}{\partial n} q^\delta
\frac{\omega\sigma_{xx}q^2}{\omega^2 + D^2 q^4}=e           
\frac{\partial\langle\sigma^{\gamma\delta}\rangle}
{\partial n} q^\delta 
 {\rm Im}\Pi^R.
\label{gb}
\end{equation}

\noindent
The density dependence of the conductivity is a measure of electron-hole 
assymetry, which is essential for Coulomb drag\cite{na}. The 
fluctuations in $\rho_D$ result from mesoscopic fluctuations in 
$\frac{\partial\sigma}{\partial n}$. 

In the absence 
of a magnetic field $\Gamma^\gamma$ is parallel to $\vec{q}$. In contrast, 
at $\nu=1/2$ the large Hall component of the conductivity leads to 
$\Gamma^\gamma$ which is approximately perpendicular to $\vec{q}$. In both 
cases the disorder-averaged conductivity is  linear in the 
density, and $\frac{\partial\sigma^{\alpha\beta}}{\partial n}
\approx{\sigma^{\alpha\beta} \over n}$. Substituting this approximation in 
Eq.~(\ref{gb}), and  using  Eqs.~(\ref{dia})  and (\ref{Ds}),
we arrive at the familiar result\cite{the4}, 

\begin{equation}
\langle\rho_D\rangle
=\frac{2\pi h}{3e^2(k_F d)^4}\left(\frac{T}{T_0}\right )^2
\ln {\left(\frac{T}{T_0}\right )}
\label{average}
\end{equation}

\noindent
with $T_0\equiv 4\pi\sigma_{xx}/\epsilon d$. Note that 
Eq. (\ref{average}) holds both at zero magnetic field and at $\nu=1/2$. 
Quantitatively, it yields very different results in the two cases, since 
the electronic longitudinal conductivity $\sigma_{xx}$ at $\nu=1/2$ is 
about three orders of magnitude smaller than at $B=0$. 

We now turn to the discussion of mesoscopic fluctuations of
$\rho_D$. The first step is the estimate of fluctuations in
$\Gamma({\bf q},\omega)$. At $\nu=1/2$ the main source of these
fluctuations is the derivative $\frac{\partial \sigma}{\partial
n}$. The rest of the parameters, i.e., the compressibility and
diffusion constant, can be approximated by their average values (as in
Eq.~(\ref{gb})), since their fluctuations are much smaller than their
average. To estimate the fluctuations of $\frac{\partial
\sigma}{\partial n}$ we express this derivative in terms of
response functions for composite fermions, which are the Fermi liquid
quasi-particles of the $\nu=1/2$ state. The conductivity matrices of
electrons and composite fermions are related by

\begin{equation}
(\hat\sigma^{cf})^{-1}=\hat\sigma^{-1}+\frac{2h}{e^2}{\hat\epsilon}
\label{cs}
\end{equation}

\noindent
with $\hat\epsilon$ being the two dimensional anti-symmetric tensor. 

On average $\sigma^{cf}$, the composite fermion conductivity matrix at 
$\nu=1/2$, is diagonal with both diagonal elements being 
$\frac{e^2}{h}g_{cf}$ (where $g_{cf}$ is the dimensionless conductance 
of composite fermions). 
In the limit of $g_{cf} \gg 1$, 
the electronic longitudinal conductivity is 
inversely proportional to $g_{cf}$,  
$\sigma_{xx}\approx \frac{e^2}{h}\frac{1}{4g_{cf}}$, and so is also the 
electronic diffusion constant. 

For a particular disorder realization the system is not 
isotropic, and $\sigma^{cf}$ is not diagonal. In the limit of 
$g_{cf}\gg1$, Eq. (\ref{cs}) leads to 

\begin{equation}
\delta {\hat \sigma}\approx 
\left(\frac{1}{4g_{cf}^2}\right){\hat\epsilon \; }
\delta {\hat \sigma}^{cf}{\; \hat\epsilon }.
\label{etocf}
\end{equation}
\noindent
where $\delta\sigma,\delta\sigma^{cf}$ are the deviations of 
$\sigma,\sigma^{cf}$ of a particular sample from their average value.
The fluctuations of $\frac{\partial \sigma}{\partial n}$ 
then stem from fluctuations of 
$\partial\sigma^{cf}/\partial n$. This derivative is taken 
with the {\it external} magnetic field $B$ kept constant. At the same time, 
the magnetic field experienced by the composite fermions, 
$\Delta B=B-2\Phi_0 n$, is not constant: when the density is 
varied, $\Delta B$ varies as well. The derivative must then be taken with 
respect to both $n$ and $\Delta B$,

\begin{equation}
\left.\frac{\partial\hat{\sigma}^{cf}}{\partial n}\right|_{B}=
\left.\frac{\partial\hat{\sigma}^{cf}}{\partial n}\right|_{\Delta B}-
\left.2\Phi_0
\frac{\partial\hat{\sigma}^{cf}}{\partial \Delta B}\right|_{n}
\label{derivatives}
\end{equation}

We now estimate the fluctuations of $\Gamma$ for a phase coherent sample 
of size $L$. The scale for the variance of the conductivity 
$\sigma^{cf}$ at $T=0$ is $\frac{e^2}{h}$. The typical magnetic 
field scale is $\Delta B\sim\Phi_0/L^2$, therefore the estimate for the 
fluctuations of $
\left.
2\Phi_0\frac{\partial\sigma^{cf}}{\partial \Delta B}\right|_{n}$ is 
$\frac{e^2}{h}L^2$. The fluctuations of 
$\left.\frac{\partial\sigma^{cf}}{\partial n}\right|_{\Delta B}$ 
are smaller by a factor of 
$g_{cf}$, and can therefore be neglected. Consequently, the variance of 
both components of $\Gamma$ is

\begin{equation}
\delta\Gamma\sim iq \frac{e}{h} \left(\frac{L}{g_{cf}}\right)^2 
{\rm Im}\Pi^R, 
\label{gamma_fluct}
\end{equation}

\noindent
Since on average 
$ \frac{\partial\langle\sigma_{xy}^{cf}\rangle}{\partial n}|_{B} 
=\frac{e^2}{2 h n}$,
the fluctuations of $\Gamma$
can be re-written as $\frac{\langle\delta\Gamma^2\rangle}
{\langle\Gamma^\gamma\rangle^2}\sim 
\left(\frac{k_F L}{g_{cf}}\right)^4$, which
 in the diffusive regime is much larger 
than unity. This by itself is a measurable conclusion: 
{\em In the diffusive 
regime in a fully coherent sample the fluctuations of the acousto-electric 
current are much larger than its average.} Equation (\ref{gamma_fluct}) 
holds as long as the 
thermal length $L_T^{cf}\equiv \sqrt{\hbar D^{cf}/T}$ and the phase 
breaking length $L_\varphi^{cf}$ are much larger than $L$.  

Having now an estimate for the fluctuations of $\Gamma$, we estimate the 
fluctuations of $\rho_D$ as: 

\begin{equation}
\delta\rho_D\sim T \int\limits_{-T}^T \frac{d\omega}{\hbar^2\omega^2}
\int\limits_{|q|\gtrsim 1/{\rm min}(L,L_\varphi^{cf})} d{\bf q}
 \ \delta\Gamma^2 
|{\cal D}({\bf q},\omega)|^2
\label{estimate-symbolic}
\end{equation}

\noindent
As commonly happens in mesoscopic fluctuations, the integral over ${\bf q}$ 
is dominated by its lower limit.
We start with the case where the phase breaking length is the largest scale
in the problem. We describe the temperature dependence of the transresistance,
and  denote the r.m.s of the fluctuations by 
$\delta\rho_D^0$.

At extremely low temperatures $T \ll (\kappa d) \hbar D/L^2 $, all the
excitations with  energies smaller than $T$ contribute and
we obtain with the help of Eqs.~(\ref{estimate-symbolic}) and (\ref{Ds})
\begin{mathletters}
\begin{equation}
\delta\rho_D^0 
\sim \frac{h}{e^2} \frac{1}{(\kappa d)^2}
\left(\frac{T}{E_T^{cf}}\right)^2,
\quad
T \ll \frac{(\kappa d)E_{T}^{cf}}{g_{cf}^2},
\label{estimate-variance1}
\end{equation} 
where $E_{T}^{cf}=\hbar D^{cf}/L^2$ is the Thouless energy for the
composite fermions.  At higher temperature $(\kappa d) \hbar D/L^2 \ll T \ll
E_{T}^{cf}$, the processes with the energy transfer $\omega > (\kappa
d) D/L^2 $ are suppressed, leading to
\begin{equation}
\delta\rho_D^0 \sim \frac{h}{e^2} \frac{1}{(\kappa d)g_{cf}^2}
\frac{T}{E_T^{cf}}, \quad 
\frac{(\kappa d)E_{T}^{cf}}{g_{cf}^2} 
\ll T \ll E_{T}^{cf}.
\label{estimate-variance2}
\end{equation}
At yet higher temperature, ${T} > E_T^{cf}$, 
$\delta \Gamma$ itself is temperature 
dependent, and it is suppressed  by a factor of  
$\sqrt{E_T^{cf}/T}$ similar to   conductance fluctuations, 
see e.g. \cite{lsf}. This yields a temperature independent result
\begin{equation}
\delta\rho_D^0 \sim \frac{h}{e^2} \frac{1}{(\kappa d)g_{cf}^2}
,\quad {T} \gg E_T^{cf}.
\label{estimate-variance3}
\end{equation}
\label{estimate-variance}
\end{mathletters}

Let us now discuss the effect of a finite phase breaking length 
$L_{\varphi}^{cf}$. As is well known for mesoscopic fluctuations,
averaging in a large sample, $L\gg L_\varphi^{cf}$, is 
carried out by summing over the statistically independent contributions 
of $(L/L_{\varphi}^{cf})^2$  patches of a size $L_\varphi^{cf}$ each.  
The contribution of each patch is given by Eqs.~(\ref{estimate-variance}), 
where the Thouless energy $E_T^{cf}$ is set to 
$\hbar/\tau_\varphi^{cf}\equiv \hbar D^{cf}/(L_\varphi^{cf})^2$. 
As a result of
the averaging $\langle\delta \rho_D^2\rangle=
(\delta\rho_D^0)^2(L_\varphi^{cf}/L)^2$. 
Using Eqs. (\ref{estimate-variance}) we obtain

\begin{eqnarray}
\langle\rho_D^2\rangle &=&\frac{h^2}{e^4}
{1 \over{g_{cf}^4(\kappa d)^2}}
\left(\frac{L_\varphi^{cf}}{L}\right)^2
{\rm min}\left[1, \alpha_1
\left(\frac{g_{cf}^2 T\tau_\varphi^{cf}}{\kappa d \hbar}\right)^2
\right]
\nonumber\\
&\times& {\rm min}\left[\alpha_3, \alpha_2
\left( T\tau_\varphi^{cf}/\hbar\right)^2
\right],
\label{final}
\end{eqnarray}
where the coefficients $\alpha_{1,2}$ are of order  unity and we were
able to calculate $\alpha_3 \approx 0.2 (32/9\pi) = 0.23$.

While the actual magnitude of the mesoscopic fluctuations depends on the
precise source of phase breaking, their temperature dependence is robust. 
All generic models of phase breaking in two dimensions lead to 
$1/\tau_{\varphi} \propto T$, so that the temperature dependence following
from Eq.~(\ref{final}) is
$\langle \delta\rho_D^2\rangle \propto 1/T$.

The dependence of the mesoscopic fluctuations on 
$g_{cf}$ requires further  specification of the model. 
Our preliminary study suggets 
that the main mechanism for the phase breaking
is the {\em quasi-elastic} scattering of composite fermions from the
thermal quasi-static fluctuations of the Chern-Simons magnetic field, 
$(2\Phi_0)^{-2}\langle B({\bf r})B({\bf r}^\prime)\rangle
= \langle \delta n ({\bf r}) \delta n ({\bf r}^\prime)\rangle
\approx \delta ({\bf r} - {\bf r}^\prime) \frac{TN(0)}{\kappa d}=
\delta ({\bf r} - {\bf r}^\prime) \frac{T\epsilon}{2\pi e^2d}$ 
(with $\epsilon$ being the bulk dielectric constant).
Due to the accumulation of the Aharonov - Bohm phase from such fluctuations, 
$(L_\varphi^{cf})^2$ is the area at which the thermal 
fluctuations of the electron number are of the order of one
$\int_{|{\bf r}|< L^{cf}_\varphi }d{\bf r}d{\bf r}^\prime
\langle n({\bf r})n({\bf r}^\prime)\rangle \sim 1$,
which result in 
\begin{equation}
T\frac{\epsilon (L^{cf}_\varphi)^2}{2\pi e^2 d} \simeq 1,
\quad \frac{\hbar}{\tau^{cf}_\varphi}\simeq \frac{g_{cf}T}{\kappa d}.
\label{tauphi}
\end{equation}
This estimate can be understood very simply: a thermal fluctuation of
a charge $e$ in one layer and $-e$ in the other layer 
over a scale $L_\varphi^{cf}$ involves
a charging energy cost of $\frac{2\pi e^2 d}{(L_\varphi^{cf})^2}$. The
energy available for that fluctuation is $T$. Balancing the two
energies determines $L_\varphi^{cf}$.
 
Substituting estimate (\ref{tauphi}) into Eq.~(\ref{final}) and using the
condition $g^{cf} \gg {\kappa d}$, we find up to a numerical constant
\begin{equation}
\langle \rho_D^2\rangle \simeq \frac{h^2}{e^4}
\left({1 \over g_{cf}^6}\right)
\left(\frac{2\pi e^2 d}{\epsilon L^2 T}\right)
\label{final2}
\end{equation}

Our estimate (\ref{tauphi}) for the phase breaking rate implies
$T\tau_\varphi\ll \hbar$.  This by no means indicates a collapse of the
Fermi liquid picture of composite fermions, since most of the phase
breaking results from scattering off the Chern-Simons field
fluctuations whose dynamics [with characteristic frequency $T/(\hbar
g_{cf})$] is very slow compared to $\tau_\varphi$, but fast compared
to the time of the experiment. Field fluctuations which are static
on the scale of the experiment time affect the mesoscopic fluctuations
only by affecting $g_{cf}$. Field fluctuations that are faster than
that scale make the potential landscape seen by the composite fermions
time dependent, and lead to a suppression of the mesoscopic
fluctuations by partial ensemble averaging.

We now use Eq.~(\ref{final2}) to estimate the value of the
fluctuations of the transresitance of a realistic sample. At $T=0.6K$ the
average drag\cite{exp5} $\langle\rho_D\rangle = 15\Omega /\Box$ with
the interlayer spacing $d=300\AA$. 
For that sample, the single layer resistance at
$\nu=1/2$ is \cite{exp5} $R=3k\Omega
/\Box$, which gives the value $g_{cf} \approx 8$. Given the sample's
size $L \simeq 100{\mu}m$, we estimate the magnitude of the
fluctuations Eq.~(\ref{final2}) as
$\delta \rho_D \approx 0.3 \Omega$. For lower
temperatures and smaller samples we expect the fluctuations to be
stronger, possibly exceeding the average. 
Increasing current supresses the mesoscopic fluctuations mostly by
heating of the electrons.

{
\narrowtext
\begin{figure}[ht]
\vspace{0.2 cm}
\epsfxsize=5 cm
\centerline{\epsfbox{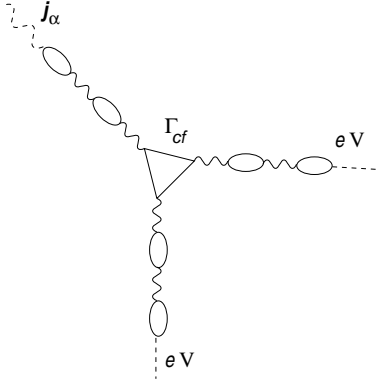}}
\vspace{0.5cm}
\caption{Electronic non-linear response 
within the composite
fermion RPA. Wavy lines are  gauge
field propagators \protect\cite{hlr}.
Bubbles are composite fermions response functions.} 
\label{1}
\end{figure}
}

We now turn to sketch the calculation, whose details will be published
separately.  Within the composite fermion random phase approximation
(RPA), $\Gamma$ is depicted diagrammatically on Fig.~\ref{1}.  The
center triangle is the non-linear response of the composite fermions
$\Gamma^{cf}$ to a driving field.  Since  composite fermions
interact with scalar and vector potentials, $\Gamma^{cf}$ is a tensor
with component $(\Gamma^{cf})^\mu_{\alpha\beta}$, where $\alpha\beta$
are the directions of the driving fields, and $\mu$ is the direction
of the induced current.

In the limit of $g_{cf}\gg 1$ the analytic expression corresponding to
the diagram on Fig.~\ref{1} is approximated by
\[
\Gamma^\gamma_{12}\approx \frac{2h^2\epsilon^{\gamma\gamma'}
}
{e^2g_{cf}}
\left[\Gamma^{cf}\right]_{34; \mu\nu}^{\gamma'}
\frac{\epsilon^{\mu\mu'}{\bf\nabla}^{\mu'}_{3}}{{\bf\nabla}^2_{3}}
\Pi^R_{31}(\omega)\frac{\epsilon^{\nu\nu'}
{\bf\nabla}^{\nu'}_{4}}{{\bf\nabla}^2_{4}}
\Pi^A_{42}(\omega),
\]
where numerical subscripts indicate spatial coordinates, we imply
integration over $r_{3,4}$, and the polarization operators are defined
in Eq.~(\ref{Ds}).  The mesoscopic fluctuations in $\Gamma$ result
from fluctuations in $\Gamma_{cf}$. Within the RPA, the latter is
approximated by the corresponing response function for non-interacting
particles: 
\begin{eqnarray} &&\left[\Gamma_{cf}\right]^{\alpha}_{12;
\mu\nu} = \int{{d\epsilon}\over{2\pi}} \left[J^{\alpha}_{12;
\mu\nu}(\omega,\epsilon) + J^{\alpha}_{21; \nu\mu}(-\omega,\epsilon)
\right],
\label{ing} \\
&&J^{\alpha}_{12; \mu\nu}(\omega,\epsilon ) =
\left(\tanh{{{\epsilon - \omega}\over{2T}}} - 
\tanh{{{\epsilon}\over{2T}}}\right) 
\nonumber\\
&&\quad
\times\left[G^R_{12}(\epsilon - \omega) - 
G^A_{12}(\epsilon - \omega)\right]j_\mu
\left[G^R(\epsilon) j^{\alpha} G^A(\epsilon) \right]_{21}j_\nu;
\nonumber
\end{eqnarray}
The further analysis of the statistics of the vertex (\ref{ing}) follows
the lines of Ref.~\cite{na}, and results in Eq.~(\ref{final}).

Closing the paper, 
we point out that
two analyses of mesoscopic fluctuations of the Coulomb drag near 
$\nu=1/2$ are of interest. First, one varies the magnetic field and measures 
the correlation function $\langle\rho_D(B)\rho_D(B+\delta B)\rangle
-\langle\rho_D(B)\rangle\langle\rho_D(B+\delta B)\rangle$, with $B$ close to
the $\nu=1/2$ value. An experimental study of the decay of this correlation 
function is a way to measure $L_\varphi^{cf}$: the characteristic magnetic 
field of the decay is $\delta B^*\sim\Phi_0/[L_\varphi^{cf}]^2$. 
Second, one varies the 
electron density in one of the layers and measures the correlation function 
$\langle\rho_D(n)\rho_D(n+\delta n)\rangle
-\langle\rho_D(n)\rangle\langle\rho_D(n+\delta n)\rangle$. Again, the decay
of this function is governed by $L_\varphi^{cf}$: 
the characteristic density change
$\delta n^*$ at which it decays is expected to correspond to half of an 
electron in a phase coherent region, i.e. $\delta n^*=1/2[L_\varphi^{cf}]^2$. 
This 
statement should hold as long as the composite fermion cyclotron radius is 
much larger than its mean free path, i.e., for 
$|\nu-1/2| < (2 g_{cf})^{-1}$. 

It is noteworthy  that near zero magnetic field 
$\delta B^*=\Phi_0/L_\varphi^2$ but $\delta n^*=k_F l/L_\varphi^2$. 
Since at $B=0$ the electrons do not carry flux tubes, the density change 
$\delta n^*$ has to be such that the chemical potential changes by 
$\hbar/\tau_\varphi$.
Thus an experimental observation of $\frac{\delta B^*}{\delta n^*}=2\Phi_0$ 
would in some sense be a verification of the attachment of flux to the Fermi 
liquid quasi-particles at and close to $\nu= 1/2$. 

I.A. is supported by Packard Foundation. AS acknowledges support of the 
US-Israel BSF, the DIP-BMBF foundation and the Victor Ehrlich chair.

\end{multicols}
\end{document}